%% file: paper.tex
\documentclass[sn-mathphys]{sn-jnl}

\usepackage[english]{babel}
\usepackage[utf8]{inputenc}
\usepackage{epstopdf}
\usepackage{graphicx}

\jyear{2022}%

\theoremstyle{thmstyleone}%
\theoremstyle{thmstyletwo}%

\theoremstyle{thmstylethree}%

\begin{document}

\title[Tricritical Dicke Triangle]{Chiral Quantum Phases and Tricriticality in a Dicke Triangle}


\author[1]{\fnm{Guo-Jing} \sur{Cheng}}

\author[2]{\fnm{Diego} \sur{Fallas Padilla}}
\author[1]{\fnm{Tao} \sur{Deng}}
\author*[1]{\fnm{Yu-Yu} \sur{Zhang}}\email{yuyuzh@cqu.edu.cn}

\author*[2]{\fnm{Han} \sur{Pu}}\email{hpu@rice.edu}

\affil[1]{\orgdiv{Department of Physics, and Chongqing Key Laboratory for strongly coupled Physics}, \orgname{Chongqing University}, \orgaddress{\city{Chongqing}, \postcode{401330}, \country{China}}}

\affil[2]{\orgdiv{Department of Physics and Astronomy, and Rice Center for Quantum Materials}, \orgname{Rice University}, \orgaddress{\city{Houston}, \state{TX}, \postcode{77251}, \country{USA}}}


\abstract{The existence of quantum tricriticality and exotic phases are found in a Dicke triangle (TDT)
where three cavities, each one containing an ensemble of three-level atoms, are connected to each other through the action of an artificial magnetic field. The conventional superradiant phase (SR) is connected to the normal phase through first- and second-order boundaries, with tricritical points located at the intersection of such boundaries.
Apart from the SR phase, a chiral superradiant (CSR) phase is found by tuning the artificial magnetic field. This phase is characterized by a nonzero photon current and its boundary presents chiral tricritical points (CTCPs). Through the study of different critical exponents, we are able to differentiate the universality class of the CTCP and TCP from that of second-order critical points, as well as find distinctive critical behavior among the two different superradiant phases. The TDT can be implemented in various systems,
including atoms in optical cavities as well as the circuit QED system, allowing the exploration of a great variety of critical manifolds.}




\maketitle

\section{Introduction}\label{sec1}

Recent efforts have been devoted to exploring
many-body quantum phases emerging in light-matter coupling systems using
different platforms such as cavity and circuit QED~\cite%
{Greentree2006,plenio,zhu2020,felicetti}, and cold atoms in optical lattices~%
\cite{bloch,duan2021,chen2021}.
The rapid development of such platforms offers high control and tunability, allowing for the exploration of richer phase diagrams with more complex critical behaviors, for example, the study of tricritical points (TCPs) and higher-order critical points (multicritical points). TCPs were originally found in He$^{3}$-He$%
^{4}$ mixtures and a simple description of their mean-field characteristics can be done using the Landau theory of phase transitions~\cite{griff}. These special points are located in the intersection of a second-order boundary and a first-order boundary, with both of them separating the same two phases~\cite%
{griff,stanley,riedel,henkel,pu,Yao2018}. Although well understood from a theoretical perspective, quantum tricriticality
is not abundant in real materials but can be found, for example, in certain metallic magnets~\cite%
{Belitz,Friedemann,Canfield,yuan2019}.

Recently, light-matter interacting systems have been proposed to realize TCPs in experiments, specifically in generalizations of the Dicke model~\cite{padilla,pu}.
The Dicke model has served historically as the cornerstone model in the description
of the interaction of light with an ensemble of identical two-level atoms~\cite{Dicke,lambert,chen2008,zhang2019}.
When the light-matter coupling strength is tuned above a critical threshold value, this system undergoes a superradiant phase transition, which has been
realized in various experimental settings~\cite{Baumann,nagy}.

Apart from interesting critical boundaries, atom-light interacting systems can be used to engineer exotic phases of matter when external fields are  incorporated. Artificial magnetic fields have been used to explore chiral
ground-state currents of interacting photons in a three-qubit loop~\cite%
{roushan}, chiral phases in a quantum Rabi triangle \cite{zhang2021}, and
fractional quantum Hall physics in the Jaynes-Cummings Hubbard lattice~\cite%
{hayward,martin2016,noh2017}. Advances in synthesizing such artificial magnetic fields have been reported in
neutral ultracold atoms~\cite{lin2009,RMP,fu} and photonic systems~\cite%
{umu,wang,Cai,roushan,bloch2012}.

Here we propose a tricritical Dicke triangle (TDT)
system as a building block for exploring all these features. The system is composed of three cavities each one containing an ensemble of three-level atoms allowing the realization of the tricritical Dicke Hamiltonian in each cavity. As a result, not only
a second-order phase transition occurs, but also a first-order transition from
the normal phase (NP) to the superradiant (SR) phase can be observed. The two types of phase boundaries meet at a conventional TCP. Interestingly, as photon hopping between neighboring cavities is permitted, new chiral
superradiant phase (CSR) and chiral tricritical points (CTCP) can be found by tuning the artificial
magnetic field, which breaks the $\mathbb{Z}_2$ and $C_3$ symmetries, causing a chiral
current of photons in the ground state. Computation of the scaling exponents shows a plethora of critical behaviors, in particular, the CTCP and conventional TCPs are found to belong to different universality classes.

\section{Single cavity tricritical Dicke model}
\begin{figure}[h]
\includegraphics[width=\textwidth]{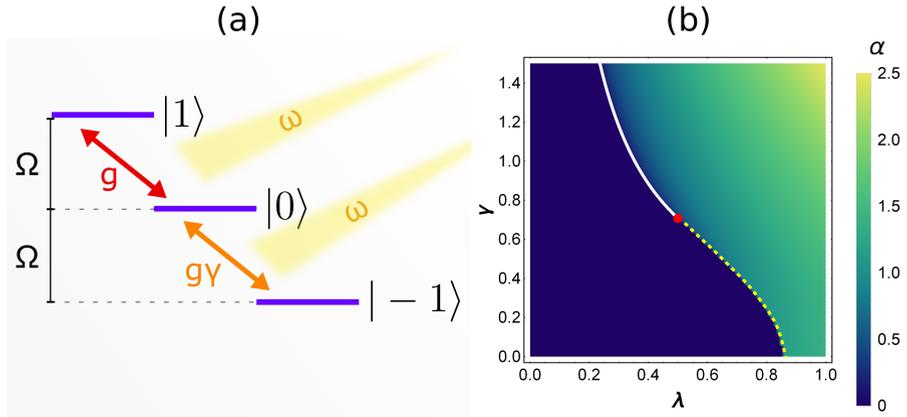}
\caption{(a) Schematic showing the atomic levels of the tricritical Dicke  model. Light with frequency $\omega$ couples the states $ \vert 1 \rangle$ and $ \vert 0 \rangle$ with interaction strength $g$, and the states $\vert 0 \rangle$ and $\vert -1\rangle$ with interaction strength $g \gamma$. (b) Phase diagram of the tricritical Dicke model in the $\lambda$-$\gamma$ plane, using $\alpha = \langle a \rangle/\sqrt{N}$ as order parameter. The white solid line, yellow dashed line and red dot indicate the second-order boundary, first-order boundary, and tricritical point, respectively.}
\label{leveld}
\end{figure}

Let us first consider a single cavity containing $N$ identical three-level atoms coupled uniformly to the cavity mode.
The Hamiltonian of this system is a  generalization of the conventional Dicke model and reads
\begin{equation}
H_{D}=\omega a^{\dagger }a+\frac{\sqrt{2}g}{\sqrt{N}}%
(a^{\dagger }+a)\sum_{k=1}^{N}d^{(k)}+\Omega \sum_{k=1}^{N}h^{(k)},
\label{TDHam}
\end{equation}%
where $a$ ($a^{\dagger }$) is the photon annihilation (creation)
operator of the single-mode cavity with frequency $\omega $, $g$ the
atom-cavity coupling strength. The dipole operator $d^{(k)}$ of the $k$-th
atom and the single-atom Hamiltonian $h^{(k)}$ are defined as
\begin{equation}
d^{(k)}=\left(
\begin{array}{ccc}
0 & 1 & 0 \\
1 & 0 & \gamma \\
0 & \gamma & 0%
\end{array}%
\right) ,\;\;\;\; h^{(k)}=\left(
\begin{array}{ccc}
1 & 0 & 0 \\
0 & 0 & 0 \\
0 & 0 & -1%
\end{array}%
\right)  \label{matrix}
\end{equation}%
where we have chosen the eigenstates of $h^{(k)}$ to be the basis states.  These states are labelled as $ \vert 1 \rangle$, $\vert 0 \rangle$ and $\vert -1 \rangle$ as shown schematically in Fig.~\ref{leveld}(a). The cavity field couples $\vert 1 \rangle$ and $\vert 0 \rangle$ as well as $\vert 0 \rangle$ and $\vert -1\rangle$, with the coupling strengths given by $g$ and $\gamma g$, respectively. The tunable dimensionless parameter $\gamma $ serves as a control parameter.

The phase diagram can be described in terms of the scaled dimensionless coupling
strength $\lambda =g/\sqrt{\Omega \omega }$ and the transition strength ratio $%
\gamma $ as shown in Fig. \ref{leveld} (b). For simplicity, we set $\omega =\Omega =1$. In the thermodynamic limit $N\rightarrow \infty$ the normal phase with $\langle a \rangle=0$ and the superradiant phase with macroscopic photon population ($\langle a \rangle \propto \sqrt{N}$), are separated by first and second-order boundaries, indicated by the yellow dashed and the white solid lines in Fig.~\ref{leveld}(b), respectively. These two types of boundaries meet at the tricritical point.  A Landau theory approach can be followed to explore the expressions for the second- and first-order phase transitions, as well as for the TCP. Here, the order parameter is given by $\alpha=\langle a \rangle / \sqrt{N}$. In terms of this order-parameter, the mean-field energy is given by:
\begin{equation}
\frac{E_{\mathtt{MF}}}{\Omega N} = \frac{1}{8 \lambda^2} \alpha^2 + \alpha d + h
\label{MFSingleCavity}
\end{equation}
where $h$ and $d$ are just the single atom operators given in Eq. \ref{matrix}, and the order parameter $\alpha$ has been rescaled by $\alpha \rightarrow \frac{2\sqrt{2}g}{\Omega}\alpha $. Due to the $Z_{2}$ symmetry of the Hamiltonian, the
mean-field ground-state energy can be expanded as a Taylor
series in terms of $\alpha ^{2}$: $E_{\mathtt{MF}}=\sum_{k=0}^{\infty
}c_{k}\alpha ^{2k}$. The coefficients $c_{k}$ are obtained through
perturbation theory, which is performed by treating $h$
as the unperturbed Hamiltonian and $d$ as the perturbation as shown in Appendix A and in Ref.~\cite{padilla}. Keeping the expansion up to order $\alpha^6$ and discarding the constant term, the mean-field energy is approximated by
\begin{equation}
\frac{E_{\mathtt{MF}}}{\Omega N}=c_{1}\alpha ^{2}+c_{2}\alpha
^{4}+c_{3}\alpha ^{6},
\end{equation}%
where $c_{1}=1 /8\lambda^{2}-\gamma ^{2}$, $c_{2}=\gamma ^{2}(\gamma ^{2}-\frac{1}{2})$, and $c_{3}=-\gamma ^{2}(1-7\gamma ^{2}+8\gamma ^{4})/4$ (see Appendix A).
An ordinary $2$nd-order critical point is obtained when $c_{1}=0$ and $%
c_{2}>0$, leading to the second-order boundary expression
\begin{equation}
\lambda_{2c} \gamma_{2c}= \frac{1}{\sqrt{8}}.
\label{2ndSingle}
\end{equation}

The tricritical point is defined by the condition $%
c_{1}=c_{2}=0$ and $c_{3}>0$~\cite{chang,griff,padilla}. For the single cavity case, the tricritical
point is located at $\gamma _{\mathtt{TCP}}=1/\sqrt{2}$ and $\lambda _{%
\mathtt{TCP}}=1/2$. When $\gamma \geq \gamma _{%
\mathtt{TCP}}$, one has $c_{1}<0$, and  $E_{\mathtt{MF}}$ goes from having a single global minimum at $\alpha=0$ for $\lambda < \lambda_{2c}$ to having two global minima
at $\alpha _{\pm }=\pm \sqrt{(-c_{2}+\sqrt{c_{2}^{2}-3c_{1}c_{3}})/3c_{3}}$ for $\lambda > \lambda_{2c}$, this behavior indicates the second-order character of the phase transition. In the case $\gamma <\gamma _{\mathtt{TCP}}$%
, $E_{\mathtt{MF}}$ has three local minima at $\alpha _{\pm }$ and $\alpha =0
$, the global minimum switches from $\alpha=0$ to $\alpha = \alpha_{\pm}$ as the phase transition is crossed, the discontinuous jump on the global minimum location characterizes the first-order boundary.  The change in the order of the transition can be clearly observed in Fig.~\ref%
{alpha} where the ground-state energy $E_{\mathtt{SR}}$ as a function of $\alpha$ is presented for different values of $\gamma$ and $\lambda$.
\begin{figure}[tbp]
\includegraphics[width=0.75\textwidth]{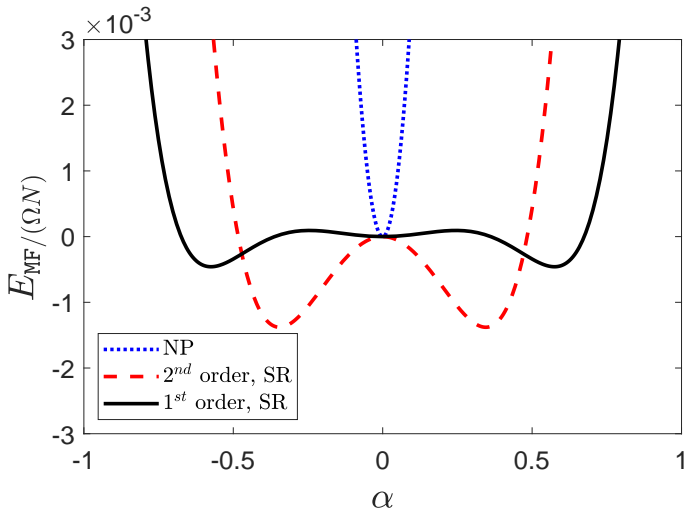}
\centering
\caption{Ground-state energy $E_{\texttt{MF}}/(\Omega N)$ as a function of the order parameter $%
\protect\alpha$ in the superradiant phase  for $\gamma<\gamma_{\texttt{TCP}}$ (black solid line), and $\gamma>\gamma_{\texttt{TCP}}$ (red
dashed line), and in the normal phase (blue dotted line), respectively. }
\label{alpha}
\end{figure}

\section{Tricritical Dicke triangle}

\begin{figure}[h]
\includegraphics[width=0.7\textwidth]{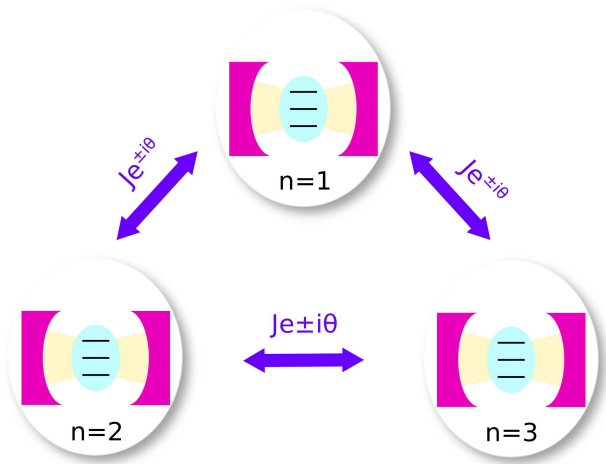}
\centering
\caption{Schematic of the TDT. Each cavity contains a three-level atom interacting with light as described in Fig.\ref{leveld} (a). Photons can hop between neighboring cavities with hopping strength $Je^{\pm i \theta}$.}
\label{TDTschematic}
\end{figure}

We now consider three such cavity systems linked by photon hopping, as schematically shown in Fig.~\ref{TDTschematic}, forming the tricritical Dicke triangle (TDT).
The TDT Hamiltonian is given by
\begin{equation}
H=\sum_{n=1}^{3}H_{D,n}+\sum_{n=1}^{3}J(e^{i\theta }a_{n}^{\dagger
}a_{n+1}+e^{-i\theta }a_{n+1}^{\dagger }a_{n}).
\end{equation}%
Here $H_{D,n}$ is the single-cavity Hamiltonian as given by Eq.~(\ref{TDHam}), where we use the sub-index $n$ to denote the $n$th cavity; $J$ is the hopping amplitude between nearest-neighbor cavities with a
phase $\theta $. The complex photon hopping amplitude means that the photons are subjected to an artificial vector potential $A(r)$ such that $\theta =\int_{r_{n}}^{r_{m}}$ $A(r)dr$ where $r_n$ and $r_m$ denote the position of the two neighboring cavities. Such an artificial vector potential or magnetic field can be achieved through temporal modulation of the
photon-hopping strength on each cavity~\cite{roushan,zhang2021}.

In analogy to the Dicke model, there exists a parity
symmetry operator $P=\Pi _{i=1}^{3}\exp \{i\pi
[a_{i}^{+}a_{i}+\sum_{k=1}^{N}(h^{(k)}+1)]\}$, which satisfies $[H,P]=0$
with eigenvalues $\pm 1$. Besides such $Z_2$ symmetry, the Hamiltonian is real when $\theta = m \pi$ ($m \in \mathbb{Z}$), and, consequently, it preserves time-reversal symmetry (TRS). When this condition is not met, the breaking of the TRS can have important implications on the behavior of photons as will be shown later.

\section{Normal Phase of TDT}

Let us first explore the normal phase (NP) of the TDT. This phase features no photon excitation just as in the single cavity case. We
employ a Schrieffer-Wolff transformation $U_{n}=\exp (\sqrt{2}g/\sqrt{N}%
S_{n})$ with an anti-Hermitian operator $S_n$ given by
\begin{equation}
S_{n}=(a_{n}^{\dagger
}+a_{n})/\Omega \sum_{k=1}^{N}\left(
\begin{array}{ccc}
0 & -1 & 0 \\
1 & 0 & -\gamma \\
0 & \gamma & 0%
\end{array}%
\right),
\end{equation}
which makes the off-diagonal terms of the single-cavity
Hamiltonian $H_{D,n}$ vanish. Neglecting higher-order terms in the thermodynamic limit $%
N\rightarrow \infty $, the transformed TDT Hamiltonian becomes
\begin{eqnarray}
H_{\mathtt{NP}}&=& \Pi_{i=1}^3 U_i^{\dagger}H \Pi_{j=1}^3 U_j \notag \\ &=&\sum_{n=1}^{3}\left(\omega a_{n}^{\dagger }a_{n}+\Omega
\sum_{k=1}^{N}h^{(k)}\right)+\sum_{n=1}^{3}J(e^{i\theta }a_{n}^{\dagger
}a_{n+1}+H.c.)  \notag \\
&&+\sum_{n=1}^3\frac{2g^2}{N\Omega }(a_{n}^{\dagger }+a_{n})^{2}\sum_{k=1}^{N}\left(
\begin{array}{ccc}
1 & 0 & 0 \\
0 & -1+\gamma ^{2} & 0 \\
0 & 0 & -\gamma ^{2}%
\end{array}%
\right) .
\end{eqnarray}%
The above Hamiltonian is diagonal in the atomic degrees of freedom, and an effective low-energy Hamiltonian can be found by projecting into the lowest energy state of the three-level atom,
\begin{eqnarray}
H_{\mathtt{NP}}^{\downarrow } &=&\sum_{n=1}^{3}(\omega -\frac{4g^{2}\gamma ^{2}}{%
\Omega })a_{n}^{\dagger }a_{n}-\frac{2g^{2}\gamma ^{2}}{\Omega }%
(a_{n}^{\dagger 2}+a_{n}^{2})  \notag \\
&&+\sum_{n=1}^{3}J(e^{i\theta }a_{n}^{\dagger }a_{n+1}+e^{-i\theta
}a_{n+1}^{\dagger }a_{n})+E_{0},
\label{effectiveNP}
\end{eqnarray}%
where the energy constant is $E_{0}=-6g^{2}\gamma ^{2}/\Omega -3N\Omega $.

Using the discrete Fourier transformation $a_{n}^{\dagger }=\sum_{q}a_q \,e^{inq}$ with the
quasi-momentum $q=0,\pm 2\pi /3$, we can rewrite the projected Hamiltonian $H_{\mathtt{NP}}^{\downarrow }$ in momentum space as
\begin{equation}
H_{\mathtt{NP}}^{\downarrow }=\sum_{q}\omega _{q}a_{q}^{\dagger }a_{q}-\frac{%
2g^{2}\gamma ^{2}}{\Omega }(a_{q}^{\dagger }a_{-q}^{\dagger
}+a_{q}a_{-q})+E_{0},
\end{equation}%
where $\omega _{q}=\omega -4g^{2}\gamma
^{2}/\Omega +2J\cos (\theta -q)$. By introducing a unitary transformation $%
S_{q}=\exp [\beta _{q}(a_{q}^{\dagger }a_{-q}^{\dagger }-a_{q}a_{-q})]$ with
a variational squeezing parameter $\beta _{q}=-\frac{1}{8}\ln \frac{\omega
_{q}+\omega _{-q}-8g^{2}\gamma ^{2}/\Omega }{\omega _{q}+\omega
_{-q}+8g^{2}\gamma ^{2}/\Omega }$, $H_{\mathtt{NP}}^{\downarrow }$ can be diagonalized and takes the
form $H_{\mathtt{NP}}^{\downarrow }=\sum_{q}\varepsilon _{q}a_{q}^{\dagger
}a_{q}+E_{g}$, where
\begin{equation}
    E_{g}=E_{0}+\frac{1}{2}\sum_{q}(\varepsilon
_{q}-\omega _{q})
\label{groundq}
\end{equation}
is the ground-state energy, and the excitation energies are
given by
\begin{equation}
\varepsilon _{q}=\frac{1}{2}[\omega _{q}-\omega _{-q}+\sqrt{%
(\omega _{q}+\omega _{-q})^{2}-64g^{4}\gamma ^{4}/\Omega ^{2}}].
\label{exciq}
\end{equation}

\begin{figure}[tbp]
\includegraphics[scale=0.6]{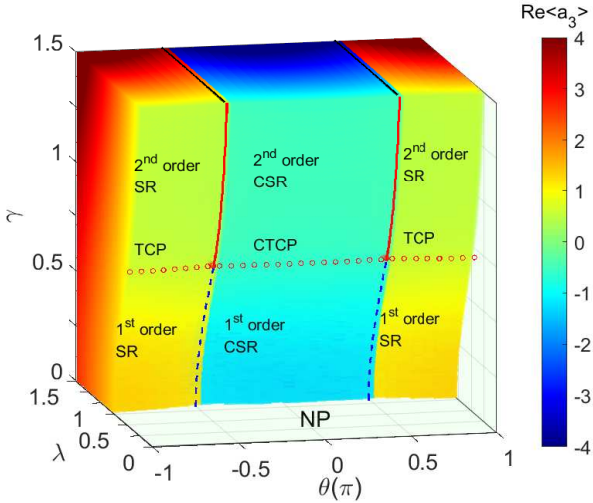}
\centering
\caption{The phase diagram of the tricritical Dicke triangle system, in
which $\protect\gamma $ is plotted as a function of the dimensionless
coupling strength $\protect\lambda $ and the hopping phase $\protect\theta $%
. The color bar represents the order parameter $A_3=Re(\langle a_{3}\rangle )$.
The red solid line is a second-order critical line while the blue dashed
line is a first-order critical line. The tricritical points are marked by red circles, in
which $\protect\gamma _{\mathtt{TCP}}=1/\protect\sqrt{2}$. The phase boundary
between the SR and CSR is denoted by solid black line, determined by $%
\vert \theta \vert =\protect\theta _{c}$. Here, we set $J/\protect%
\omega =0.1$.}
\label{phase}
\end{figure}

A second-order phase transition occurs when the gap between the first excited state and the ground state vanishes, then, the condition $\varepsilon_q=0$ can be used to determine the location of these boundaries, leading to the critical values
\begin{equation}
\lambda _{2c}\gamma _{2c}=\sqrt{\frac{1+4J/\omega \cos q\cos \theta
+4J^{2}/\omega ^{2}\cos (\theta -q)\cos (\theta +q)}{8(1+2J/\omega \cos
\theta \cos q)}}.  \label{critical}
\end{equation}%
Note that if $\theta$ and $J/\omega$ are fixed, the expression above signals a $\gamma$-$\lambda$ second-order line, but since $\theta$ will be taken as an additional control parameter that can vary, Eq.~(\ref{critical}) refers, in general, to a second-order surface in the three dimensional parameter space spanned by $\gamma$, $\lambda$ and $\theta$, as shown in Fig.~\ref{phase}. Moreover, Eq.~(\ref{critical}) describes two different second-order boundaries, one for $q=0$ and the other for $q=\pm 2\pi/3$, as discussed in the following sections, each of these $q$-values is associated with a different superradiant phase. Additionally, note that Eq.~(\ref{critical}) reduces to Eq.~(\ref{2ndSingle}) for the single cavity case if we take the limit of no hopping between cavities $J=0$, which is expected.

\section{Superradiant phases of TDT}

As the coupling strength increases to $\lambda >\lambda _{\mathtt{2c}}$, the
number of photons in each cavity becomes proportional to $N$. To capture the superradiant physics, the bosonic
operators are shifted as $a_{n}\rightarrow a_{n}+\sqrt{N}\alpha _{n},$ $%
a_{n}^{\dagger }\rightarrow a_{n}^{\dagger }+\sqrt{N}\alpha _{n}^{\ast }$
with the complex displacement parameter $\alpha _{n}=A_{n}+iB_{n}$. Note that in the NP $\alpha_n=0$. The
transformed Hamiltonian becomes
\begin{eqnarray}\label{SRHamiltonian}
H_{\texttt{SR}}&=&\sum_{n=1}^{3}\omega \tilde{a}_{n}^{\dagger }\tilde{a}_{n}+\sum_{k=1}^{N}\sqrt{\frac{2%
}{N}}g(\tilde{a}_{n}^{\dagger }+\tilde{a}_{n})d^{(k)}+\Omega
\sum_{k=1}^{N}h^{(k)} \nonumber\\
&&+\sum_{n=1}^{3}J\tilde{a}_{n}^{\dagger }(e^{i\theta
}\tilde{a}_{n+1}+e^{-i\theta }\tilde{a}_{n-1})+H_{l}+E_{\mathtt{SR}},
\end{eqnarray}
where the linear term is $H_{l}=\sum_{n=1}^{3}\omega \sqrt{N}%
(\tilde{a}_{n}^{\dagger }\alpha _{n}+\tilde{a}_{n}\alpha _{n}^{\ast })+\sqrt{N}%
J[\tilde{a}_{n}^{\dagger }(e^{i\theta }\alpha _{n+1}+e^{-i\theta }\alpha _{n-1})+h.c]$. The ground-state energy is expressed as
\begin{eqnarray}
\frac{E_{\mathtt{SR}}}{N} &=&\sum_{n=1}^{3}2\sqrt{2}gA_{n}d+\Omega h+\omega
(A_{n}^{2}+B_{n}^{2})  \notag  \label{csrenergy} \\
&+&JA_{n}[\cos \theta (A_{n+1}+A_{n-1})-\sin \theta (B_{n+1}-B_{n-1})]
\notag \\
&+&JB_{n}[\cos \theta (B_{n+1}+B_{n-1})+\sin \theta (A_{n+1}-A_{n-1})].
\notag \\
&&
\label{MFenergy}
\end{eqnarray}%
with $d$ and $h$ being the three-level operators in Eq.~(\ref{matrix}).
The mean-field values $A_n$ and $B_n$ used to characterize the different phases are found by minimizing the energy given in Eq.~(\ref{MFenergy}), where two types of superradiant phases can be identified depending on whether $\vert \theta \vert$ is greater or lower than $\theta_c$ (see below).  The complete phase diagram of the TDT is presented in Fig.~\ref{phase}. Note that although there are six order parameters, namely, $A_n$ and $B_n$ with $n=1,2,3$, the value $A_3$ has been chosen to describe the phase diagram. Nonetheless, relations between all six-order parameters are provided in the following sections.

\subsection{Conventional superradiant phase (SR)}
In the SR phase $\alpha _{n}=A_n$ is real and non-zero, and is the same for
all three cavities, $\alpha _{n}=\alpha _{n\pm1}$. Then, each cavity behaves as an independent tricritical Dicke model. For a given value of $\theta$, the
boundary between the SR and NP phases is split into a second-order line (the red solid line) and a first-order critical line (the blue dashed line) as shown in Fig.~\ref{phase}. The two lines merge together in the TCP (represented with the red dots). In the three dimensional parameter space shown in Fig.~\ref{phase}, the TCPs form a line.
The second-order phase boundary is consistent with the
analytical expression $\gamma_{\mathtt{2c}}$-$\lambda_{\mathtt{2c}}$ in Eq. (%
\ref{critical}) with the momentum $q=0$.

\begin{figure}[tbp]
\includegraphics[width=0.75\textwidth]{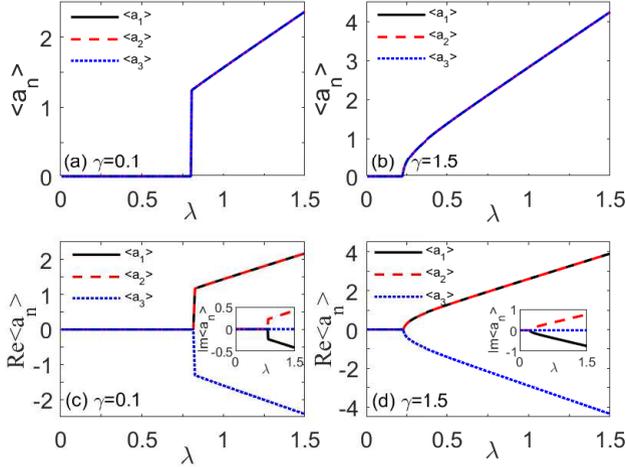}
\centering
\caption{ The order parameter $\langle a_{n}\rangle =\alpha_n$ as a function of the
dimensionless coupling strength $\protect\lambda$ across the NP-SR phase
transition of (a) first-order $\protect\gamma =0.1$, and (b) second-order $\protect\gamma =1.5$. $\theta$ is kept fixed at $2\pi/3>\theta_c$. The real
part of $\langle a_{n}\rangle$ across the NP-CSR phase
transition of (c) first-order $\protect\gamma =0.1$, and (d) second-order $\protect\gamma =1.5$. $\theta$ is kept fixed at $\pi/3<\theta_c$. The inset corresponds to the
imaginary part $B_n=\mathtt{Im}(\langle a_{n}\rangle)$ of the order parameter.}
\label{order}
\end{figure}

Figure \ref{order} (a)(b) shows the order parameter $\alpha_n$ for
the NP-SR phase transition as a function of $%
\lambda $. For a small value of the atomic transition ratio $\gamma=0.1$, $%
\alpha_n$ is zero in the NP, and increases with an abrupt jump in
the SR phase in Fig.~\ref{order} (a), indicating a first-order phase
transition. Since the transition between the middle and the
upper state of the three-level atom dominates for a small $\gamma$, $\langle h_n\rangle>0$ increases abruptly as well across the first-order transition (see Appendix).
However, for $\gamma= 1.5$ in Fig.~\ref{order} (b), $\langle a_{n}\rangle$ changes smoothly
from the NP to SR phase, exhibiting a second-order phase
transition. Note that, in the SR, the ground state is two-fold degenerate as the configurations break the $Z_2$ symmetry. In Figs.~\ref{phase} and \ref{order}, one of the degenerate configurations is chosen, the other one is simply obtained by changing the sign of the order parameter.

A perturbation theory analysis can be done in a similar fashion as for the single cavity case. Similarly, the mean-field energy for the SR can be approximated by
\begin{equation}  \label{ESR}
\frac{E_{\mathtt{SR}}}{3\Omega N}=c_{1}\alpha ^{2}+c_{2}\alpha
^{4}+c_{3}\alpha ^{6},
\end{equation}%
with $c_1 = \frac{\omega' \Omega}{8 g^2} $, where $\omega^{\prime }=\omega +2J\cos \theta $. $c_2$ and $c_3$ have the same form as in the single cavity case. Consequently, the second order boundary expression for the SR is given by
\begin{equation}
\lambda _{2c}^{\mathtt{SR}}\gamma _{2c}^{\mathtt{SR}} =\frac{1}{2\sqrt{2}}%
\sqrt{1+2J/\omega \cos \theta },
\end{equation}
which is consistent with Eq.~(\ref{critical}) for $q=0$. The tricritical point is located at $\gamma _{\mathtt{TCP}}=1/\sqrt{2}$ and $\lambda _{\mathtt{TCP}}=\sqrt{1+2J/\omega \cos \theta }/2$. Note that, as expected, both of these results reduce to the single cavity case if the limit $J=0$ is taken.

\subsection{Chiral superradiant phase (CSR)}
In the CSR phase, $\alpha _{n}$
is complex and depends on $n$. Minimization of the mean-field energy in Eq.~(\ref{csrenergy}) yields
\begin{eqnarray}
A_n \neq A_{n+1} = A_{n-1}, \quad B_n =0, \quad B_{n+1} = B_{n-1}.
\end{eqnarray}
Since the solutions above break both the $Z_{2}$ and $C_3$ symmetries, the ground state in the CSR is six-fold degenerate. For a clear presentation of results we choose the particular solution $A_3 \neq A_1=A_2$, $B_3=0$, $B_2=-B_1$.

In Fig.~\ref{order} (c)(d) the order parameter $\alpha_n$ is shown for the CSR phase transition. As observed in both panels, the order parameter is site-dependent, contrary to the SR case. However, there is still a change in the order of the transition depending on the $\gamma$ value. Consequently, just as in the SR phase, there are chiral tricritical points (CTCPs) in the CSR phase as observed in Fig.~\ref{phase}.

To investigate the phase boundaries in the CSR phase, we start with a particular solution $A_{3}=A$, $A_{1}=A_{2}=\widetilde{A}$ and $B_{3}=0$. Similar to the SR case,
the mean-field ground-state energy can be written as a Taylor series in
terms of $A^{2}$ and $\widetilde{A}^{2}$%
\begin{eqnarray}  \label{energy2}
\frac{E_{\mathtt{CSR}}}{\Omega N} &=&2(\omega _{\mathtt{CSR}}+J^{\prime
}-\gamma ^{2})\widetilde{A}^{2}+(\omega _{\mathtt{CSR}}-\gamma ^{2})A^{2}
\notag  \label{Ecsr} \\
&&+4J^{\prime }A\widetilde{A}+c_{2}(A^{4}+2\widetilde{A}^{4}),
\end{eqnarray}%
where $\omega _{\mathtt{CSR}}=[\omega -2J^{2}\sin ^{2}\theta /(\omega -J\cos
\theta )]\Omega /8g^{2}$, $J^{\prime }=\Omega /8g^{2}(J\cos \theta +\frac{%
J^{2}\sin ^{2}\theta }{\omega -J\cos \theta })$ and the coefficient $%
c_{2}=\gamma ^{2}(\gamma ^{2}-\frac{1}{2})$. By minimizing the energy using $%
\partial E/\partial A=0$ and $\partial E/\partial \widetilde{A}=0$, the
expression for the second-order boundary in the CSR can be found to be
\begin{eqnarray}\label{lambda2}
\lambda _{2c}^{\mathtt{CSR}}\gamma _{2c}^{\mathtt{CSR}} &=&\frac{1}{2\sqrt{2}%
}\sqrt{\frac{1-2J/\omega \cos \theta +J^{2}/\omega ^{2}(\cos ^{2}\theta
-3\sin ^{2}\theta )}{1-J/\omega \cos \theta }},
\end{eqnarray}%
which is consistent with Eq.~(\ref%
{critical}) when choosing  $q=\pm 2\pi /3$. As expected, the critical line $\gamma$-$\lambda $ of the second-order
NP-CSR transition in Fig.~\ref{phase} fits well with the analytical
solutions $\lambda _{2c}^{\mathtt{CSR}}$. The CTCP is located at $\gamma _{%
\mathtt{CTCP}}=1/\sqrt{2} $ as a consequence of setting $c_{2}=0$ in Eq.(\ref%
{energy2}). By substituting $\gamma _{\mathtt{CTCP%
}}$ into Eq. (\ref{lambda2}), $\lambda _{\mathtt{CTCP}}$ can be determined.

If $\gamma$ and $\lambda$ are fixed inside the SR phase, and $\theta$ is varied until entering the CSR the order parameter changes discontinuously. Thus, the phase transition between the two superradiant phases is of first-order and indicated by the solid black line in Fig.~\ref{phase}.
Right at the boundary between the SR and
CSR phases, conditions $B_n=0$ and $A_{n-1}=A_{n+1}=\pm A_n$ need to be satisfied. From Eq. (\ref{csrenergy}), this implies $J \cos%
\theta+J^2 \sin^2\theta/(\omega-J \cos\theta)=0$, which leads to the critical hopping phase that separates the SR and CSR
\begin{equation}
\theta _{c}=\cos ^{-1}\left(-\frac{2J}{\sqrt{8J^{2}+\omega ^{2}}+\omega }\right).
\end{equation}%
The entire superradiance region is split into the CSR phase regime for $%
\vert \theta \vert \leq \theta_c $ and the SR phase regime for $\vert \theta \vert >\theta_c$.

\begin{figure}[tbp]
\includegraphics[trim=100 50 50 30,scale=0.18]{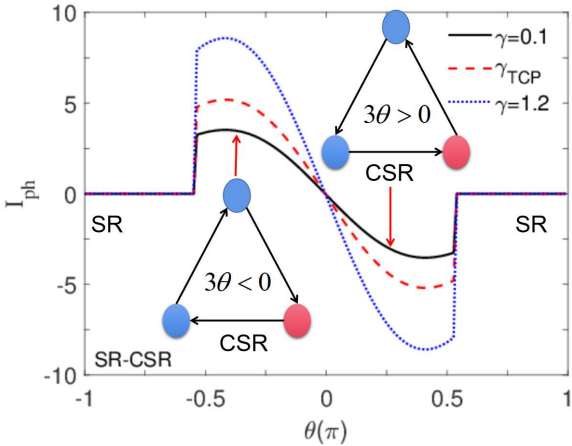}
\centering
\caption{Chiral photon current $I_{\mathtt{ph}}$ as a function of the
hopping phase $\protect\theta$ for $\protect\gamma=0.1$, $\protect\gamma_{%
\mathtt{TCP}}$, and $1.2$, respectively. Schematic of the  mean photon population in the
three cavities in the CSR region,
two cavities with the blue color have the same excitation of photons and are
different from the third cavity denoted in red. Here $%
\protect\lambda=1$ and $J/\protect\omega=0.1$.}
\label{Iph}
\end{figure}

To characterize further the chirality in the CSR phase, we analyze the ground-state
current of photons in the closed loop of three cavities. Similar to the
continuity equation in classical systems, the photon current operator can be
explicitly defined as
\begin{eqnarray}
I_{\mathrm{ph}} &=&i\left[ (a_{1}^{\dagger }a_{2}+a_{2}^{\dagger
}a_{3}+a_{3}^{\dagger }a_{1})-h.c.\right]
\end{eqnarray}
Fig.~\ref{Iph} shows the photon current in the ground state for $%
\lambda>\lambda_{\mathtt{2c}}$ in the SR and CSR phases. By varying the
effective magnetic flux $\theta$, a discontinuous jump of $I_{\mathrm{ph}}$
is observed at the critical hopping phase $\pm\theta_c$. $I_{\mathrm{ph}}$
goes from zero in the SR phase, to a  non-zero value in
the CSR phase and changes its sign depending on the phase $\theta$. Then, varying $\theta$ changes the orientation of the photons circulating in the loop from clockwise to anticlockwise, a signature of
the chiral phase. The ground-state current of photons is associated with the
nonuniform excitation of photons in three cavities, which is induced by the
magnetic flux.

\section{Critical behavior}

Second-order phase transitions are characterized by their scaling exponents in the vicinity of the transition. Here, we explore the critical behavior of the total photon number $N_{\mathtt{ph}}=\sum_{n=1}^3
\langle a_n^\dagger a_n\rangle$ near three important regions: the second-order critical boundary, the TCP
and the CTCP, in order to classify the universality of each of these critical manifolds.

\begin{figure}[tbp]
\includegraphics[width=0.8\textwidth]{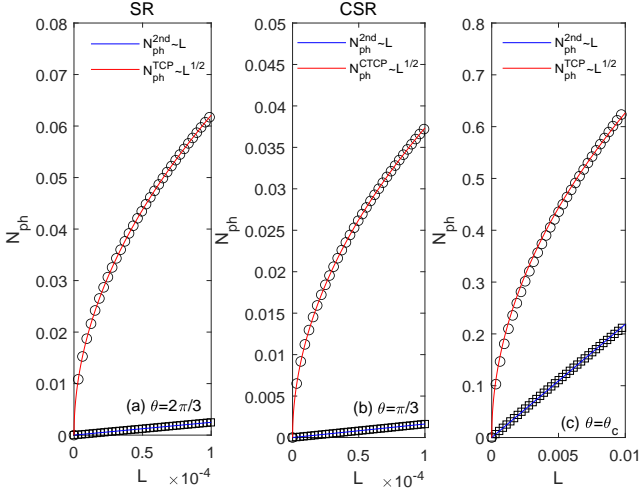}
\centering
\caption{Mean photons in three cavities $N_{ph}$ as a function of the distance $L$ between a point inside the corresponding superradiant phase and a critical point located in the boundary with the normal phase for the CSR phase (a), the SR phase (b), and a triple point (c). Conventional second-order critical points ($\gamma=0.9$) are represented by open squares, while TCP's and CTCP's are represented with open circles. The corresponding fitting lines are listed.}
\label{exponents}
\end{figure}

First, we consider a point ($\gamma$, $\lambda$) in the SR region and
close to the second-order critical line for fixed $\theta=2\pi/3$. A line through
this point is perpendicular to the critical line and intercepts the critical
line at a second-order critical point ($\gamma_{\mathtt{2c}}^{\mathtt{SR}}$%
, $\lambda_{\mathtt{2c}}^{\mathtt{SR}}$)~\cite{pu}. Around the critical point,
the photon number scales like $N_{\mathtt{ph}}\propto L^{\beta}$, where $%
L\equiv \sqrt{(\lambda-\lambda_{\mathtt{2c}}^{\mathtt{SR}})^2+(\gamma-\gamma_{%
\mathtt{2c}}^{\mathtt{SR}})^2}$ is the distance between the point and the second-order critical point. Fig.~\ref{exponents} (a)  displays $N_{\mathtt{ph}}$ at ($\gamma$, $\lambda$) as a function of $L$.  The critical
exponent for this transition is $\beta=1$, consequently, $N_{\mathtt{ph}}^{\mathtt{2nd}}\propto L$.
However, if the perpendicular line through ($\gamma$, $\lambda$) intercepts
the critical line at the TCP ($\gamma_{\mathtt{TCP}}$, $\lambda_{\mathtt{TCP}}$%
), one has a different scaling
\begin{eqnarray}
N_{\mathtt{ph}}^{\mathtt{TCP}}\propto L^{1/2}
\end{eqnarray}
which gives a critical exponent $1/2$ for the TCP. This illustrates that the TCP
belongs to a different universality class in comparison to the conventional second-order critical points.

Fig.~\ref{exponents} (b) show the scaling behaviors for the CSR phase at $\theta=\pi/3$. The mean photon number in one of the sites is different from that in the other two cavities due to the break of the $C_3$ symmetry. However, we found that the photon number in each cavity has the same scaling behavior. Similar to the SR phase, the critical exponent for the $2$nd-order
critical point ($\gamma_{\mathtt{2c}}^{\mathtt{CSR}}$, $\lambda_{\mathtt{2c}%
}^{\mathtt{CSR}}$) is obtained to be $1$.  The scaling function
at the CTCP ($\gamma_{\mathtt{CTCP}}$, $\lambda_{\mathtt{CTCP}}$) is found to be
\begin{eqnarray}
N_{\mathtt{ph}}^{\mathtt{CTCP}}\propto L^{1/2}.
\end{eqnarray}
This indicates that all tricritical points, regardless whether they are TCP or CTCP, have the same scaling exponent for photon numbers.

Finally, Fig.~\ref{exponents} (c) shows the scaling exponents at $\theta=\theta_c$ at the critical line. This line is special since it represents the line of triple points at which three phases (SR, CSR, and NP) coexist. The scaling along this line shows the same behavior, which is expected, as both SR and CSR have the same scaling exponents.

The exponent of $N_{\mathtt{ph}}$ ($\beta$ )is useful to distinguish between ordinary critical points and tricritical ones, nonetheless, it does not signal any differences between the SR and CSR phase transitions, which is unexpected as both phases have very distinct features, the scaling behavior of other quantities could be useful for further characterization of the critical behavior. To this end, let us examine other critical exponents such as the scaling of the excitation energy.
The effective low-energy Hamiltonian on Eq.~(\ref{effectiveNP}) has a quadratic form in the $a_n$ operators, consequently, a Bogoliubov transformation can be performed to diagonalize the Hamiltonian in the form
 \begin{equation}
 H_{\mathtt{NP}}^{\downarrow} = \sum_{q} \varepsilon_q a^{\dagger}_q a_q + E_g,
 \label{quadeffect}
 \end{equation}
with $q=0,\pm 2\pi/3$, $E_g$ being the ground state energy given in Eq.~(\ref{groundq}), $\varepsilon_q$ being the excitation energies given in Eq.~(\ref{exciq}), and $a_q$($a_q^{\dagger}$) being a new set of annihilation (creation) operators obtained through the Bogoliubov transformation. Precisely at the critical points, the lowest of the set of excitation energies $\{\varepsilon_q\}$ vanishes, and we denote the lowest excitation energy by $\varepsilon_1$. consequently, we expect that around the critical point this quantity behaves as $\varepsilon_1 \propto L^\eta$.

\begin{figure}[tbp]
\includegraphics[width=0.75\textwidth]{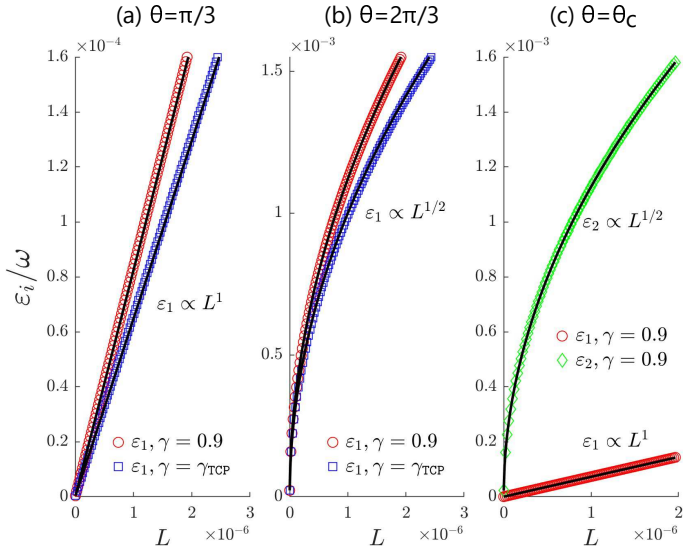}
\centering
\caption{Excitation energies $\varepsilon_i$ as a function of the distance $L$ between a point in the normal phase and critical point located in the boundary with the CSR phase (a), the SR phase (b), or a triple point (c). $\varepsilon_1$ denotes the lowest excitation energy, and $\varepsilon_2$ is the second-lowest. Both conventional second-order critical points ($\gamma=0.9$) and tricritical points (TCP's and CTCP's) with $\gamma=\gamma_{\mathtt{TCP}}$ are explored. The fitting lines are shown on each panel. Here, $J/\omega=0.01$.}
\label{excitation}
\end{figure}

Since Eq.~(\ref{quadeffect}) is only valid in the normal phase, $L$ in this case is the distance between a point in the normal phase and the critical point. In Fig.~\ref{excitation}, the scaling of $\varepsilon_1$ as a function of $L$ is shown for both the SR and CSR phases. In the SR we find that $\varepsilon_1 = \varepsilon_{q=0}$, while in the CSR $\varepsilon_1=\varepsilon_{q=\pm 2\pi/3}$.
The exponent $\eta$ is found to be 1 for the CSR while it has a value of $1/2$ for the SR. This means that the excitation behavior on the onset of the phase boundaries is different between such phases. However, this exponent does not seem to be responsive to the order of the critical point as tricritical points follow the same behavior as conventional critical points.

Moreover, we note that at a triple point ($\theta=\theta_c$) two excitation energies vanish (denoted by $\varepsilon_1$ and $\varepsilon_2$) instead of just one, as a sign of the coexistence of both superradiant phases at this point This behavior is illustrated in Fig.~\ref{excitation} (c). Consistently, the two $\eta$ exponents are found to be $1/2$ and $1$ representing the SR and CSR phase transitions, respectively.

A study of $\varepsilon_1$ for points inside the superradiant region could retrieve some interesting behavior as described in Refs.~\cite{padilla2022understanding,zhao2022frustrated}. However, an effective Hamiltonian of the form of Eq.~(\ref{effectiveNP}) is not easily obtainable for the TDT in the superradiant regions. Nonetheless, it seems that the complementary use of $\eta$ and $\beta$ exponents already allow us to characterize the critical behavior of the different points and boundaries in the system, illustrating the great variety of features that can be explored using the TDT.

\section{Conclusion}
A different transition ratio between atomic levels and the incorporation of an artificial magnetic field make the tricritical Dicke triangle an ideal platform for studying the interplay between higher-order critical points and chiral phases of matter. Two different superradiant phases can be found by tuning the phase $\theta$ of the photon hopping amplitude, and both of them can be accessed from the normal phase through first- and second-order transitions, as well as tricritical points. The scaling behavior of the excitation energy indicates that the NP-SR and the NP-CSR transitions belong to different universality classes; while the scaling behavior of the photon mean-field population elucidates a different universality between tricritical and ordinary critical points, making evident the richness of critical manifolds in the system.
Our study opens intriguing avenues for exploring quantum tricriticality and rich phases in a single light-matter interacting platform.

\backmatter

\bmhead{Acknowledgments}

YYZ was supported by NSFC under Grant No.12075040 and No. 12147102,
Chongqing NSF under Grants No. cstc2020jcyj-msxmX0890, and Fundamental
Research Funds for the Central Universities Grant No. 2021CDJQY-007. HP
acknowledges support from the US NSF and the Welch Foundatioin (Grant No.
C-1669).

\begin{appendices}

\section{ Coefficients $c_{k}$'s by perturbation theory}
We apply perturbation theory to obtain the coefficients $c_{k}$'s in Eq.
(\ref{ESR}). The mean-field Hamiltonian in the SR phase is given by
\begin{eqnarray} \label{Ea}
\frac{E_{\mathtt{SR}}}{3N\Omega } &=&\frac{\omega +2Jcos\theta }{\Omega }%
\alpha ^{2}+H_{a}, \\
H_{a} &=&D+h.
\end{eqnarray}%
with $\alpha ^{\prime }=2\sqrt{2}g\alpha /\Omega $ and $D=\alpha^{\prime }d$%
. $h$ is treated as the unperturbed Hamiltonian, which gives the eigenstates
$\vert \varepsilon _{i}\rangle $ ($i=1,2,3$) with the eigenvalues $\varepsilon
_{1}=-1$, $\varepsilon _{2}=0$ and $\varepsilon _{3}=1$. And $d$ is the
perturbation term, the coefficients $c_{k}$ are obtained by the perturbation
expansion up to $(2k)$ order. The ground-state wave function can be
expanded as
\begin{eqnarray}
\vert \varphi \rangle &=&\vert \varepsilon _{1}\rangle +\sum\limits_{m\neq 1}\frac{%
\vert m\rangle \langle m\vert }{E-\varepsilon _{m}}D\vert \varphi \rangle  \notag \\
&=&\vert \varepsilon _{1}\rangle +G(E)D \vert \varphi \rangle ,
\end{eqnarray}%
where $G(E)=\sum\limits_{m\neq 1} \vert m\rangle \langle m \vert /(E-\varepsilon _{m})$
and $H_{a}\vert \varphi \rangle =E\vert \varphi \rangle $. This means that the wave function can be found through iteration as:
\begin{eqnarray}
\vert \varphi \rangle &=& \vert \varepsilon _{1}\rangle +G(E)D \vert n\rangle
+G(E)DG(E)D\vert n\rangle  \notag \\
&&+G(E)DG(E)DG(E)D \vert n\rangle +...
\end{eqnarray}
From $D \vert \varphi \rangle =(E-\varepsilon _{1}) \vert \varphi \rangle $, we obtain
the ground-state energy
\begin{equation}
E-\varepsilon _{1}=\langle \varepsilon _{1} \vert D \vert \varphi \rangle .
\end{equation}%
By substituting the wave function into the equation above, the ground-state energy
is given by
\begin{eqnarray}
E &=&\varepsilon _{1}+\langle \varepsilon _{1} \vert D \vert \varepsilon _{1}\rangle
+\langle \varepsilon _{1}\vert DG(E)D\vert \varepsilon _{1}\rangle  \notag \\
&&+\langle \varepsilon _{1}\vert DG(E)DG(E)D\vert \varepsilon _{1}\rangle +...
\end{eqnarray}
Clearly, the zero-th energy correction is $E^{(0)}=\varepsilon _{1}$. Since $\langle
\varepsilon _{1}\vert D \vert \varepsilon _{1}\rangle $ is zero due to the symmetry of the Hamiltonian, the first non-zero correction is the second-order one
\begin{eqnarray}
E^{(2)} &=&\varepsilon _{1}+\langle \varepsilon _{1} \vert DG(E)D \vert \varepsilon
_{1}\rangle  \notag \\
&=&\varepsilon _{1}+\frac{\vert D_{12} \vert ^{2}}{E^{(0)}-\varepsilon _{2}} =-1-\alpha
^{\prime 2}\gamma ^{2}.
\end{eqnarray}%
The fourth-order correction of the ground-state energy is

\begin{eqnarray}
E^{(4)} &=&\varepsilon _{1}+\langle \varepsilon _{1}\vert DG(E)D \vert \varepsilon
_{1}\rangle +\langle \varepsilon _{1}\vert DG(E)DG(E)DG(E)D \vert \varepsilon
_{1}\rangle   \notag \\
&=&\varepsilon _{1}+\alpha ^{\prime 2}\frac{\vert d_{12} \vert ^{2}}{%
E^{(2)}-\varepsilon _{2}} +\alpha ^{\prime 4}\sum\limits_{m\neq 1}\sum_{n\neq 1}\sum_{k\neq
1}\langle \varepsilon _{1} \vert d\frac{\vert m\rangle \langle m \vert }{E^{(0)}-\varepsilon
_{m}}d\frac{ \vert n\rangle \langle n \vert }{E^{(0)}-\varepsilon _{n}}d\frac{ \vert k\rangle
\langle k\vert }{E^{(0)}-\varepsilon _{k}}d\vert \varepsilon _{1}\rangle   \notag \\
&=&-1+\frac{\alpha ^{\prime 2}\gamma ^{2}}{-1-\alpha ^{\prime 2}\gamma ^{2}}-%
\frac{1}{2}\alpha ^{\prime 4}\gamma ^{2}.
\end{eqnarray}

Since $\alpha $ is small around the critical point, the above energy can
be approximated by
\begin{eqnarray}
E^{(4)} &=&-1-\alpha ^{\prime 2}\gamma ^{2}+\gamma ^{2}(\gamma ^{2}-\frac{1}{%
2})\alpha ^{\prime 4}.
\end{eqnarray}
The sixth-order correction of the energy is given by

\begin{eqnarray}
E^{(6)} &=&\varepsilon _{1}+\langle \varepsilon _{1} \vert DG(E)D \vert \varepsilon
_{1}\rangle +\langle \varepsilon _{1}\vert DG(E)DG(E)DG(E)D \vert \varepsilon
_{1}\rangle   \notag \\
&&+\langle \varepsilon _{1}\vert DG(E)DG(E)DG(E)DG(E)DG(E)D\vert \varepsilon
_{1}\rangle   \notag \\
&=&\varepsilon _{1}+\alpha ^{\prime 2}\frac{ \vert d_{12}\vert ^{2}}{%
E^{(4)}-\varepsilon _{2}}+\alpha ^{\prime 4}\sum\limits_{m\neq 1}\sum_{n\neq 1}\sum_{k\neq
1}\langle \varepsilon _{1} \vert d\frac{\vert m\rangle \langle m\vert }{E^{(2)}-\varepsilon
_{m}}d\frac{\vert n\rangle \langle n \vert }{E^{(2)}-\varepsilon _{n}}d\frac{ \vert k\rangle
\langle k \vert }{E^{(2)}-\varepsilon _{k}}d \vert \varepsilon _{1}\rangle   \notag \\
&&+\alpha ^{\prime 6}\sum\limits_{m\neq 1}\sum_{n\neq 1}\sum_{k\neq
1}\sum_{i\neq 1}\sum_{j\neq 1} \frac{\langle \varepsilon _{1} \vert m\rangle
d_{mn} }{E^{(0)}-\varepsilon _{m}}\frac{d_{nk}}{%
E^{(0)}-\varepsilon _{n}}\frac{d_{ki} }{E^{(0)}-\varepsilon
_{k}}\frac{d_{ij} }{E^{(0)}-\varepsilon _{i}}\frac{
\langle j\vert \varepsilon _{1}\rangle }{E^{(0)}-\varepsilon _{j}}   \notag \\
&=&\varepsilon _{1}+\frac{\alpha ^{\prime 2}\gamma ^{2}}{E^{(4)}-\varepsilon
_{2}}+\alpha ^{\prime 4}\frac{\vert d_{12}\vert ^{2}}{(E^{(2)}-\varepsilon _{2})^{2}}%
\frac{\vert d_{23}\vert ^{2}}{E^{(2)}-\varepsilon _{3}}+\alpha ^{\prime 6}\frac{\vert d_{12} \vert ^{3}}{(E^{(0)}-\varepsilon _{2})^{3}}%
\frac{\vert d_{23}\vert ^{3}}{(E^{(0)}-\varepsilon _{3})^{2}}.\notag \\
\end{eqnarray}
Then, the ground-state energy up to the sixth-order in perturbation can be
approximately given as a power series in terms of $\alpha^2 $%
\begin{eqnarray}
E^{(6)} &=&-1-\alpha ^{\prime 2}\gamma ^{2}+\gamma ^{2}(\gamma ^{2}-\frac{1}{%
2})\alpha ^{\prime 4}  \notag \\
&&-\frac{1}{4}\gamma ^{2}(8\gamma ^{4}-7\gamma ^{2}+1)\alpha ^{\prime 6}.
\end{eqnarray}

\begin{figure}[tbp]
\includegraphics[scale=0.63]{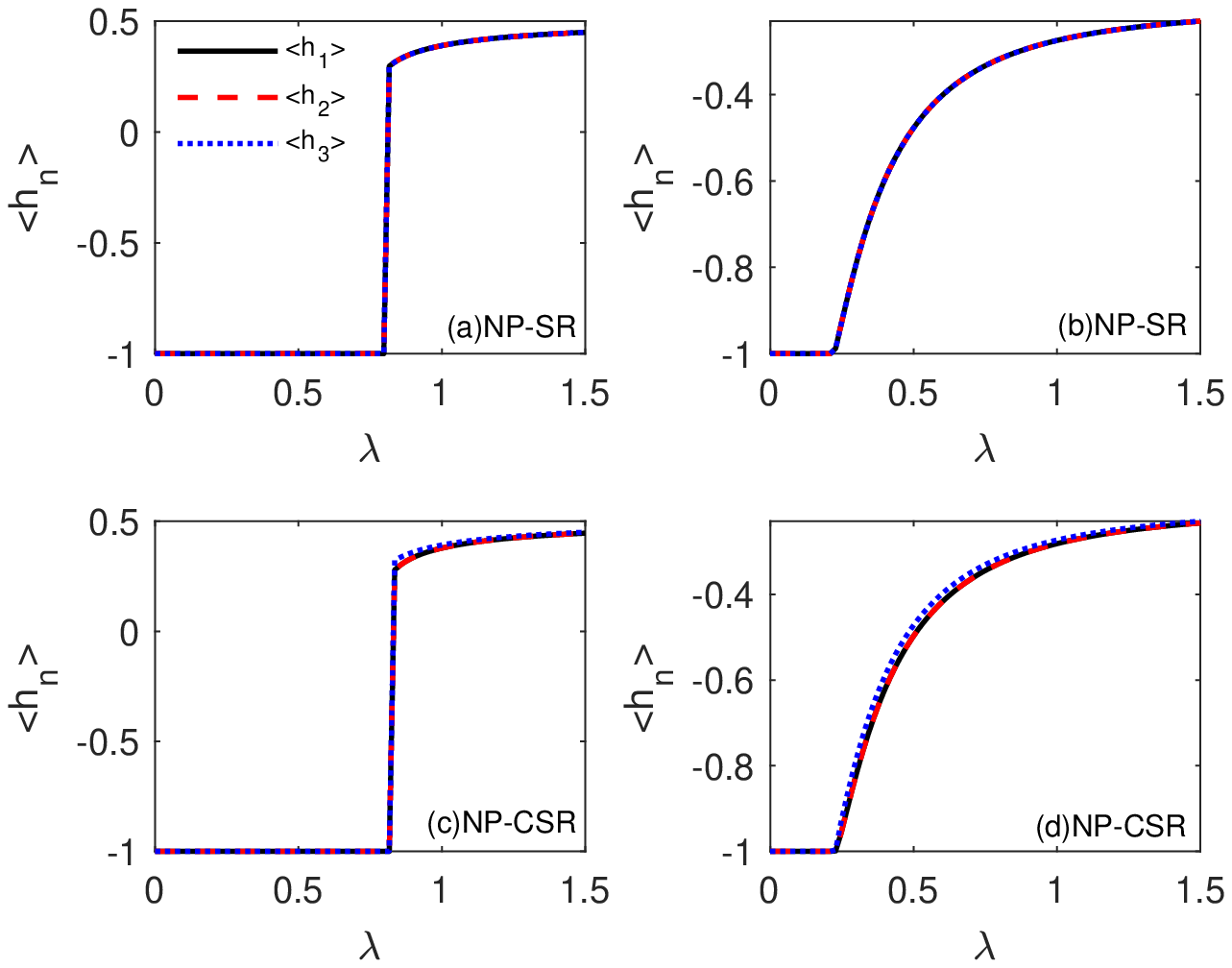}
\caption{$\langle h_{n}\rangle $ of a single atom in the $n$-th cavity as a function of the
dimensionless coupling strength $\protect\lambda$ across the NP-SR phase
transition of (a) first-order $\protect\gamma =0.1$, and (b) second-order $\protect\gamma =1.5$. $\theta$ is kept fixed at $2\pi/3>\theta_c$. $\langle h_{n}\rangle$ s a function of the
dimensionless coupling strength $\protect\lambda$ across the NP-CSR phase
transition of (c) first-order $\protect\gamma =0.1$, and (d) second-order $\protect\gamma =1.5$. $\theta$ is kept fixed at $\pi/3<\theta_c$.}
\label{atom}
\end{figure}

The expected value of $\langle h_{n}\rangle $ for a single atom in the $n$-th
cavity is calculated by minimizing the energy in Eq. (\ref{Ea}). Fig.~\ref{atom}
shows $\langle h_{n}\rangle $ in the three cavities for the first- and
second-order phase transitions from the NP to SR and CSR phases. In the NP phase,
the atom stays in the down state with $\langle h_{n}\rangle=-1$. For the
first-order phase transition (panels (a) and (c)), $\langle h_{n}\rangle $ exhibits an abrupt
jump from $-1$ to $\langle h_{n}\rangle>0 $. In contrast, $\langle
h_{n}\rangle $ increases smoothly from $-1$ across the second-order phase
transition with $\gamma=1.5$ as show in panels (b) and (d).

\end{appendices}


\bibliography{refs}
\input{paper.bbl}


\end{document}

%% file: paper.bbl

%% file: paper.bbl
\begin{thebibliography}{39}
\ifx \bisbn   \undefined \def \bisbn  #1{ISBN #1}\fi
\ifx \binits  \undefined \def \binits#1{#1}\fi
\ifx \bauthor  \undefined \def \bauthor#1{#1}\fi
\ifx \batitle  \undefined \def \batitle#1{#1}\fi
\ifx \bjtitle  \undefined \def \bjtitle#1{#1}\fi
\ifx \bvolume  \undefined \def \bvolume#1{\textbf{#1}}\fi
\ifx \byear  \undefined \def \byear#1{#1}\fi
\ifx \bissue  \undefined \def \bissue#1{#1}\fi
\ifx \bfpage  \undefined \def \bfpage#1{#1}\fi
\ifx \blpage  \undefined \def \blpage #1{#1}\fi
\ifx \burl  \undefined \def \burl#1{\textsf{#1}}\fi
\ifx \doiurl  \undefined \def \doiurl#1{\url{https://doi.org/#1}}\fi
\ifx \betal  \undefined \def \betal{\textit{et al.}}\fi
\ifx \binstitute  \undefined \def \binstitute#1{#1}\fi
\ifx \binstitutionaled  \undefined \def \binstitutionaled#1{#1}\fi
\ifx \bctitle  \undefined \def \bctitle#1{#1}\fi
\ifx \beditor  \undefined \def \beditor#1{#1}\fi
\ifx \bpublisher  \undefined \def \bpublisher#1{#1}\fi
\ifx \bbtitle  \undefined \def \bbtitle#1{#1}\fi
\ifx \bedition  \undefined \def \bedition#1{#1}\fi
\ifx \bseriesno  \undefined \def \bseriesno#1{#1}\fi
\ifx \blocation  \undefined \def \blocation#1{#1}\fi
\ifx \bsertitle  \undefined \def \bsertitle#1{#1}\fi
\ifx \bsnm \undefined \def \bsnm#1{#1}\fi
\ifx \bsuffix \undefined \def \bsuffix#1{#1}\fi
\ifx \bparticle \undefined \def \bparticle#1{#1}\fi
\ifx \barticle \undefined \def \barticle#1{#1}\fi
\bibcommenthead
\ifx \bconfdate \undefined \def \bconfdate #1{#1}\fi
\ifx \botherref \undefined \def \botherref #1{#1}\fi
\ifx \url \undefined \def \url#1{\textsf{#1}}\fi
\ifx \bchapter \undefined \def \bchapter#1{#1}\fi
\ifx \bbook \undefined \def \bbook#1{#1}\fi
\ifx \bcomment \undefined \def \bcomment#1{#1}\fi
\ifx \oauthor \undefined \def \oauthor#1{#1}\fi
\ifx \citeauthoryear \undefined \def \citeauthoryear#1{#1}\fi
\ifx \endbibitem  \undefined \def \endbibitem {}\fi
\ifx \bconflocation  \undefined \def \bconflocation#1{#1}\fi
\ifx \arxivurl  \undefined \def \arxivurl#1{\textsf{#1}}\fi
\csname PreBibitemsHook\endcsname

\bibitem{Greentree2006}
\begin{barticle}
\bauthor{\bsnm{Greentree}, \binits{A.D.}},
\bauthor{\bsnm{Tahan}, \binits{C.}},
\bauthor{\bsnm{Cole}, \binits{J.H.}},
\bauthor{\bsnm{Hollenberg}, \binits{L.C.}}:
\batitle{Quantum phase transitions of light}.
\bjtitle{Nature Physics}
\bvolume{2}(\bissue{12}),
\bfpage{856}--\blpage{861}
(\byear{2006})
\end{barticle}
\endbibitem

\bibitem{plenio}
\begin{barticle}
\bauthor{\bsnm{Hartmann}, \binits{M.J.}},
\bauthor{\bsnm{Brandao}, \binits{F.G.}},
\bauthor{\bsnm{Plenio}, \binits{M.B.}}:
\batitle{Strongly interacting polaritons in coupled arrays of cavities}.
\bjtitle{Nature Physics}
\bvolume{2}(\bissue{12}),
\bfpage{849}--\blpage{855}
(\byear{2006})
\end{barticle}
\endbibitem

\bibitem{zhu2020}
\begin{barticle}
\bauthor{\bsnm{Zhu}, \binits{C.}},
\bauthor{\bsnm{Ping}, \binits{L.}},
\bauthor{\bsnm{Yang}, \binits{Y.}},
\bauthor{\bsnm{Agarwal}, \binits{G.S.}}:
\batitle{Squeezed light induced symmetry breaking superradiant phase
  transition}.
\bjtitle{Physical Review Letters}
\bvolume{124}(\bissue{7}),
\bfpage{073602}
(\byear{2020})
\end{barticle}
\endbibitem

\bibitem{felicetti}
\begin{barticle}
\bauthor{\bsnm{Felicetti}, \binits{S.}},
\bauthor{\bsnm{Le~Boit{\'e}}, \binits{A.}}:
\batitle{Universal spectral features of ultrastrongly coupled systems}.
\bjtitle{Physical Review Letters}
\bvolume{124}(\bissue{4}),
\bfpage{040404}
(\byear{2020})
\end{barticle}
\endbibitem

\bibitem{bloch}
\begin{barticle}
\bauthor{\bsnm{Bloch}, \binits{I.}},
\bauthor{\bsnm{Dalibard}, \binits{J.}},
\bauthor{\bsnm{Zwerger}, \binits{W.}}:
\batitle{Many-body physics with ultracold gases}.
\bjtitle{Reviews of modern physics}
\bvolume{80}(\bissue{3}),
\bfpage{885}
(\byear{2008})
\end{barticle}
\endbibitem

\bibitem{duan2021}
\begin{barticle}
\bauthor{\bsnm{Cai}, \binits{M.-L.}},
\bauthor{\bsnm{Liu}, \binits{Z.-D.}},
\bauthor{\bsnm{Zhao}, \binits{W.-D.}},
\bauthor{\bsnm{Wu}, \binits{Y.-K.}},
\bauthor{\bsnm{Mei}, \binits{Q.-X.}},
\bauthor{\bsnm{Jiang}, \binits{Y.}},
\bauthor{\bsnm{He}, \binits{L.}},
\bauthor{\bsnm{Zhang}, \binits{X.}},
\bauthor{\bsnm{Zhou}, \binits{Z.-C.}},
\bauthor{\bsnm{Duan}, \binits{L.-M.}}:
\batitle{Observation of a quantum phase transition in the quantum rabi model
  with a single trapped ion}.
\bjtitle{Nature communications}
\bvolume{12}(\bissue{1}),
\bfpage{1}--\blpage{8}
(\byear{2021})
\end{barticle}
\endbibitem

\bibitem{chen2021}
\begin{barticle}
\bauthor{\bsnm{Chen}, \binits{X.}},
\bauthor{\bsnm{Wu}, \binits{Z.}},
\bauthor{\bsnm{Jiang}, \binits{M.}},
\bauthor{\bsnm{L{\"u}}, \binits{X.-Y.}},
\bauthor{\bsnm{Peng}, \binits{X.}},
\bauthor{\bsnm{Du}, \binits{J.}}:
\batitle{Experimental quantum simulation of superradiant phase transition
  beyond no-go theorem via antisqueezing}.
\bjtitle{Nature communications}
\bvolume{12}(\bissue{1}),
\bfpage{1}--\blpage{8}
(\byear{2021})
\end{barticle}
\endbibitem

\bibitem{griff}
\begin{barticle}
\bauthor{\bsnm{Griffiths}, \binits{R.B.}}:
\batitle{Phase diagrams and higher-order critical points}.
\bjtitle{Phys. Rev. B}
\bvolume{12},
\bfpage{345}--\blpage{355}
(\byear{1975})
\end{barticle}
\endbibitem

\bibitem{stanley}
\begin{barticle}
\bauthor{\bsnm{Chang}, \binits{T.S.}},
\bauthor{\bsnm{Hankey}, \binits{A.}},
\bauthor{\bsnm{Stanley}, \binits{H.E.}}:
\batitle{Generalized scaling hypothesis in multicomponent systems. i.
  classification of critical points by order and scaling at tricritical
  points}.
\bjtitle{Phys. Rev. B}
\bvolume{8},
\bfpage{346}--\blpage{364}
(\byear{1973})
\end{barticle}
\endbibitem

\bibitem{riedel}
\begin{barticle}
\bauthor{\bsnm{Riedel}, \binits{E.K.}}:
\batitle{Scaling approach to tricritical phase transitions}.
\bjtitle{Phys. Rev. Lett.}
\bvolume{28},
\bfpage{675}--\blpage{678}
(\byear{1972})
\end{barticle}
\endbibitem

\bibitem{henkel}
\begin{botherref}
\oauthor{\bsnm{Henkel}, \binits{M.}}:
Conformal invariance and critical phenomena
(2013)
\end{botherref}
\endbibitem

\bibitem{pu}
\begin{barticle}
\bauthor{\bsnm{Xu}, \binits{Y.}},
\bauthor{\bsnm{Pu}, \binits{H.}}:
\batitle{Emergent universality in a quantum tricritical dicke model}.
\bjtitle{Phys. Rev. Lett.}
\bvolume{122},
\bfpage{193201}
(\byear{2019})
\end{barticle}
\endbibitem

\bibitem{Yao2018}
\begin{barticle}
\bauthor{\bsnm{Yin}, \binits{S.}},
\bauthor{\bsnm{Jian}, \binits{S.-K.}},
\bauthor{\bsnm{Yao}, \binits{H.}}:
\batitle{Chiral tricritical point: A new universality class in dirac systems}.
\bjtitle{Phys. Rev. Lett.}
\bvolume{120},
\bfpage{215702}
(\byear{2018})
\end{barticle}
\endbibitem

\bibitem{Belitz}
\begin{barticle}
\bauthor{\bsnm{Belitz}, \binits{D.}},
\bauthor{\bsnm{Kirkpatrick}, \binits{T.R.}}:
\batitle{Quantum triple point and quantum critical end points in metallic
  magnets}.
\bjtitle{Phys. Rev. Lett.}
\bvolume{119},
\bfpage{267202}
(\byear{2017})
\end{barticle}
\endbibitem

\bibitem{Friedemann}
\begin{barticle}
\bauthor{\bsnm{Friedemann}, \binits{S.}},
\bauthor{\bsnm{Duncan}, \binits{W.J.}},
\bauthor{\bsnm{Hirschberger}, \binits{M.}},
\bauthor{\bsnm{Bauer}, \binits{T.}},
\bauthor{\bsnm{Kuchler}, \binits{R.}},
\bauthor{\bsnm{Neubauer}, \binits{A.}},
\bauthor{\bsnm{Brando}, \binits{M.}},
\bauthor{\bsnm{Pfleiderer}, \binits{C.}},
\bauthor{\bsnm{Grosche}, \binits{F.M.}}:
\batitle{Quantum tricritical points in nbfe2}.
\bjtitle{Nature Physics}
\bvolume{14},
\bfpage{62}--\blpage{67}
(\byear{2017})
\end{barticle}
\endbibitem

\bibitem{Canfield}
\begin{barticle}
\bauthor{\bsnm{Kaluarachchi}, \binits{U.S.}},
\bauthor{\bsnm{Taufour}, \binits{V.}},
\bauthor{\bsnm{Bud'ko}, \binits{S.L.}},
\bauthor{\bsnm{Canfield}, \binits{P.C.}}:
\batitle{Quantum tricritical point in the temperature-pressure-magnetic field
  phase diagram of ${\mathrm{cetige}}_{3}$}.
\bjtitle{Phys. Rev. B}
\bvolume{97},
\bfpage{045139}
(\byear{2018})
\end{barticle}
\endbibitem

\bibitem{yuan2019}
\begin{barticle}
\bauthor{\bsnm{Wu}, \binits{F.}},
\bauthor{\bsnm{Guo}, \binits{C.Y.}},
\bauthor{\bsnm{Chen}, \binits{Y.}},
\bauthor{\bsnm{Su}, \binits{H.}},
\bauthor{\bsnm{Wang}, \binits{A.}},
\bauthor{\bsnm{Smidman}, \binits{M.}},
\bauthor{\bsnm{Yuan}, \binits{H.Q.}}:
\batitle{Magnetic field induced antiferromagnetic tricritical points in
  $\mathrm{Ce}{}_{2}\mathrm{Sb}$ and $\mathrm{Ce}{}_{2}\mathrm{Bi}$}.
\bjtitle{Phys. Rev. B}
\bvolume{99},
\bfpage{064419}
(\byear{2019})
\end{barticle}
\endbibitem

\bibitem{padilla}
\begin{barticle}
\bauthor{\bsnm{Xu}, \binits{Y.}},
\bauthor{\bsnm{Fallas~Padilla}, \binits{D.}},
\bauthor{\bsnm{Pu}, \binits{H.}}:
\batitle{Multicriticality and quantum fluctuation in a generalized dicke
  model}.
\bjtitle{Phys. Rev. A}
\bvolume{104},
\bfpage{043708}
(\byear{2021})
\end{barticle}
\endbibitem

\bibitem{Dicke}
\begin{barticle}
\bauthor{\bsnm{Dicke}, \binits{R.H.}}:
\batitle{Coherence in spontaneous radiation processes}.
\bjtitle{Phys. Rev.}
\bvolume{93},
\bfpage{99}--\blpage{110}
(\byear{1954})
\end{barticle}
\endbibitem

\bibitem{lambert}
\begin{barticle}
\bauthor{\bsnm{Lambert}, \binits{N.}},
\bauthor{\bsnm{Emary}, \binits{C.}},
\bauthor{\bsnm{Brandes}, \binits{T.}}:
\batitle{Entanglement and the phase transition in single-mode superradiance}.
\bjtitle{Phys. Rev. Lett.}
\bvolume{92},
\bfpage{073602}
(\byear{2004})
\end{barticle}
\endbibitem

\bibitem{chen2008}
\begin{barticle}
\bauthor{\bsnm{Chen}, \binits{Q.-H.}},
\bauthor{\bsnm{Zhang}, \binits{Y.-Y.}},
\bauthor{\bsnm{Liu}, \binits{T.}},
\bauthor{\bsnm{Wang}, \binits{K.-L.}}:
\batitle{Numerically exact solution to the finite-size dicke model}.
\bjtitle{Phys. Rev. A}
\bvolume{78},
\bfpage{051801}
(\byear{2008})
\end{barticle}
\endbibitem

\bibitem{zhang2019}
\begin{barticle}
\bauthor{\bsnm{Chen}, \binits{X.-Y.}},
\bauthor{\bsnm{Zhang}, \binits{Y.-Y.}}:
\batitle{Finite-size scaling analysis in the two-photon dicke model}.
\bjtitle{Phys. Rev. A}
\bvolume{97},
\bfpage{053821}
(\byear{2018})
\end{barticle}
\endbibitem

\bibitem{Baumann}
\begin{barticle}
\bauthor{\bsnm{Kristian~Baumann}, \binits{F.} \bsuffix{Christine~Guerlin}},
\bauthor{\bsnm{Brennecke}, \binits{T.E.}}:
\batitle{Dicke quantum phase transition with a superfluid gas in an optical
  cavity}.
\bjtitle{Nature}
\bvolume{464},
\bfpage{1301}--\blpage{6}
(\byear{2010})
\end{barticle}
\endbibitem

\bibitem{nagy}
\begin{barticle}
\bauthor{\bsnm{Nagy}, \binits{D.}},
\bauthor{\bsnm{K\'onya}, \binits{G.}},
\bauthor{\bsnm{Szirmai}, \binits{G.}},
\bauthor{\bsnm{Domokos}, \binits{P.}}:
\batitle{Dicke-model phase transition in the quantum motion of a bose-einstein
  condensate in an optical cavity}.
\bjtitle{Phys. Rev. Lett.}
\bvolume{104},
\bfpage{130401}
(\byear{2010})
\end{barticle}
\endbibitem

\bibitem{roushan}
\begin{barticle}
\bauthor{\bsnm{Roushan}, \binits{P.}},
\bauthor{\bsnm{Neill}, \binits{C.}},
\bauthor{\bsnm{Megrant}, \binits{A.}},
\bauthor{\bsnm{Chen}, \binits{Y.}},
\bauthor{\bsnm{Babbush}, \binits{R.}},
\bauthor{\bsnm{Barends}, \binits{R.}},
\bauthor{\bsnm{Campbell}, \binits{B.}},
\bauthor{\bsnm{Chen}, \binits{Z.}},
\bauthor{\bsnm{Chiaro}, \binits{B.}},
\bauthor{\bsnm{Dunsworth}, \binits{A.}}, \betal:
\batitle{Chiral ground-state currents of interacting photons in a synthetic
  magnetic field}.
\bjtitle{Nature Physics}
\bvolume{13}(\bissue{2}),
\bfpage{146}--\blpage{151}
(\byear{2017})
\end{barticle}
\endbibitem

\bibitem{zhang2021}
\begin{barticle}
\bauthor{\bsnm{Zhang}, \binits{Y.-Y.}},
\bauthor{\bsnm{Hu}, \binits{Z.-X.}},
\bauthor{\bsnm{Fu}, \binits{L.}},
\bauthor{\bsnm{Luo}, \binits{H.-G.}},
\bauthor{\bsnm{Pu}, \binits{H.}},
\bauthor{\bsnm{Zhang}, \binits{X.-F.}}, \betal:
\batitle{Quantum phases in a quantum rabi triangle}.
\bjtitle{Physical Review Letters}
\bvolume{127}(\bissue{6}),
\bfpage{063602}
(\byear{2021})
\end{barticle}
\endbibitem

\bibitem{hayward}
\begin{barticle}
\bauthor{\bsnm{Hayward}, \binits{A.L.}},
\bauthor{\bsnm{Martin}, \binits{A.M.}},
\bauthor{\bsnm{Greentree}, \binits{A.D.}}:
\batitle{Fractional quantum hall physics in jaynes-cummings-hubbard lattices}.
\bjtitle{Physical Review Letters}
\bvolume{108}(\bissue{22}),
\bfpage{223602}
(\byear{2012})
\end{barticle}
\endbibitem

\bibitem{martin2016}
\begin{barticle}
\bauthor{\bsnm{Hayward}, \binits{A.L.}},
\bauthor{\bsnm{Martin}, \binits{A.M.}}:
\batitle{Superfluid-mott transitions and vortices in the
  jaynes-cummings-hubbard lattices with time-reversal-symmetry breaking}.
\bjtitle{Physical Review A}
\bvolume{93}(\bissue{2}),
\bfpage{023828}
(\byear{2016})
\end{barticle}
\endbibitem

\bibitem{noh2017}
\begin{barticle}
\bauthor{\bsnm{Noh}, \binits{C.}},
\bauthor{\bsnm{Angelakis}, \binits{D.G.}}:
\batitle{Quantum simulations and many-body physics with light}.
\bjtitle{Reports on Progress in Physics}
\bvolume{80}(\bissue{1}),
\bfpage{016401}
(\byear{2016})
\end{barticle}
\endbibitem

\bibitem{lin2009}
\begin{barticle}
\bauthor{\bsnm{Lin}, \binits{Y.-J.}},
\bauthor{\bsnm{Compton}, \binits{R.L.}},
\bauthor{\bsnm{Jim{\'e}nez-Garc{\'\i}a}, \binits{K.}},
\bauthor{\bsnm{Porto}, \binits{J.V.}},
\bauthor{\bsnm{Spielman}, \binits{I.B.}}:
\batitle{Synthetic magnetic fields for ultracold neutral atoms}.
\bjtitle{Nature}
\bvolume{462}(\bissue{7273}),
\bfpage{628}--\blpage{632}
(\byear{2009})
\end{barticle}
\endbibitem

\bibitem{RMP}
\begin{barticle}
\bauthor{\bsnm{Dalibard}, \binits{J.}},
\bauthor{\bsnm{Gerbier}, \binits{F.}},
\bauthor{\bparticle{Juzeli\ifmmode~\bar{u}\else} \bsnm{\={u}\fi{}nas},
  \binits{G.}},
\bauthor{\bsnm{\"Ohberg}, \binits{P.}}:
\batitle{Colloquium: Artificial gauge potentials for neutral atoms}.
\bjtitle{Rev. Mod. Phys.}
\bvolume{83},
\bfpage{1523}--\blpage{1543}
(\byear{2011})
\end{barticle}
\endbibitem

\bibitem{fu}
\begin{barticle}
\bauthor{\bsnm{Cao}, \binits{H.}},
\bauthor{\bsnm{Wang}, \binits{Q.}},
\bauthor{\bsnm{Fu}, \binits{L.-B.}}:
\batitle{Interaction effects in a quantum simulation of classical magnetism
  with artificial gauge potential}.
\bjtitle{Phys. Rev. A}
\bvolume{89},
\bfpage{013610}
(\byear{2014})
\end{barticle}
\endbibitem

\bibitem{umu}
\begin{barticle}
\bauthor{\bparticle{Umucal\ifmmode \imath \else~\i} \bsnm{\fi{}lar},
  \binits{R.O.}},
\bauthor{\bsnm{Carusotto}, \binits{I.}}:
\batitle{Fractional quantum hall states of photons in an array of dissipative
  coupled cavities}.
\bjtitle{Phys. Rev. Lett.}
\bvolume{108},
\bfpage{206809}
(\byear{2012})
\end{barticle}
\endbibitem

\bibitem{wang}
\begin{barticle}
\bauthor{\bsnm{Wang}, \binits{D.-W.}},
\bauthor{\bsnm{Cai}, \binits{H.}},
\bauthor{\bsnm{Liu}, \binits{R.-B.}},
\bauthor{\bsnm{Scully}, \binits{M.O.}}:
\batitle{Mesoscopic superposition states generated by synthetic spin-orbit
  interaction in fock-state lattices}.
\bjtitle{Phys. Rev. Lett.}
\bvolume{116},
\bfpage{220502}
(\byear{2016})
\end{barticle}
\endbibitem

\bibitem{Cai}
\begin{barticle}
\bauthor{\bsnm{Han~Cai}, \binits{D.-W.W.}}:
\batitle{Topological phases of quantized light}.
\bjtitle{National Science Review}
\bvolume{8},
\bfpage{196}
(\byear{2021})
\end{barticle}
\endbibitem

\bibitem{bloch2012}
\begin{barticle}
\bauthor{\bsnm{Bloch}, \binits{I.}},
\bauthor{\bsnm{Dalibard}, \binits{J.}},
\bauthor{\bsnm{Nascimbene}, \binits{S.}}:
\batitle{Quantum simulations with ultracold quantum gases}.
\bjtitle{Nature Physics}
\bvolume{8}(\bissue{4}),
\bfpage{267}--\blpage{276}
(\byear{2012})
\end{barticle}
\endbibitem

\bibitem{chang}
\begin{barticle}
\bauthor{\bsnm{Chang}, \binits{T.S.}},
\bauthor{\bsnm{Hankey}, \binits{A.}},
\bauthor{\bsnm{Stanley}, \binits{H.E.}}:
\batitle{Generalized scaling hypothesis in multicomponent systems. i.
  classification of critical points by order and scaling at tricritical
  points}.
\bjtitle{Phys. Rev. B}
\bvolume{8},
\bfpage{346}--\blpage{364}
(\byear{1973})
\end{barticle}
\endbibitem

\bibitem{padilla2022understanding}
\begin{botherref}
\oauthor{\bsnm{Padilla}, \binits{D.F.}},
\oauthor{\bsnm{Pu}, \binits{H.}},
\oauthor{\bsnm{Cheng}, \binits{G.-J.}},
\oauthor{\bsnm{Zhang}, \binits{Y.-Y.}}:
Understanding the quantum rabi ring using analogies to quantum magnetism.
arXiv preprint arXiv:2207.07763
(2022)
\end{botherref}
\endbibitem

\bibitem{zhao2022frustrated}
\begin{barticle}
\bauthor{\bsnm{Zhao}, \binits{J.}},
\bauthor{\bsnm{Hwang}, \binits{M.-J.}}:
\batitle{Frustrated superradiant phase transition}.
\bjtitle{Physical Review Letters}
\bvolume{128}(\bissue{16}),
\bfpage{163601}
(\byear{2022})
\end{barticle}
\endbibitem

\end{thebibliography}
